\documentclass[useAMS,usenatbib]{mn2e}
\usepackage{graphicx}
\usepackage{natbib}
\usepackage{amssymb}
\usepackage{setspace}
\usepackage{amssymb}
\usepackage{epsfig,color}
\usepackage{multicol}

\long\def\symbolfootnote[#1]#2{\begingroup%
\def\thefootnote{\fnsymbol{footnote}}\footnote[#1]{#2}\endgroup} 

\newcommand{\cc}{~cm$^{-3}$}   % cm^-3
\title[Clump mass function and early MC evolution: I.]{Clump mass function at an early stage of molecular cloud evolution: I. A statistical approach}
\author[Donkov, Veltchev \& Klessen]
  {Sava Donkov$^{1\,\star}$, Todor V.~Veltchev$^{2,3}$, and Ralf S.~Klessen$^3$ \\
  $^1$Department of Applied Physics, Technical University, 8 Kliment Ohridski Blvd., 1000 Sofia, Bulgaria \\
  $^2$University of Sofia, Faculty of Physics, 5 James Bourchier Blvd., 1164 Sofia, Bulgaria\\
  $^3$Zentrum f\"ur Astronomie der Universit\"at Heidelberg, Institute of Theoretical Astrophysics, Albert-\"Uberle-Str. 2, 69120 Heidelberg, Germany}

\date{Submitted 2011 Xxxxx XX}

\pagerange{\pageref{firstpage}--\pageref{lastpage}} \pubyear{2011}

\begin{document}

\label{firstpage}

\maketitle

\begin{abstract}
We derive the mass function of condensations (clumps) which were formed through a turbulent cascade over a range of spatial scales $L\le20$~pc during early, predominantly turbulent evolution of a molecular cloud. The approach rests upon the assumption of a statistical clump mass-density relationship $n\propto m^x$ with a scale dependence of the exponent $x$ obtained from equipartition relations between various forms of energy of clumps. The derived clump mass function (ClMF) could be represented by series of 2 or 3 power laws, depending on the chosen equipartition relation, the velocity scaling index and the type of turbulent forcing. The high-mass ClMF exhibits an average slope $\Gamma\simeq-1$, typical for fractal clouds, whereas its intermediate-mass part is shallower or flattened, in agreement with some observational studies.  
\end{abstract}

\begin{keywords}
ISM: clouds - ISM: structure - turbulence - methods: statistical
\end{keywords}

\section{Introduction} 
Understanding of the formation and the evolution of molecular clouds (MC) is of key significance in the  theory of star formation. The early MC evolution, prior to subsequent processes of active star formation, allows for simplified physical modelling due to lack of feedback from the emerging stars. Recent numerical simulations shed light on this epoch \citep{VS_ea_07, Henne_ea_08, Baner_ea_09}. Its characteristic stages could be summarized as follows: i) convergent flows in the warm neutral medium lead to local compressions and non-linear instabilities; ii) turbulent domains (clouds) of cold molecular gas form in the dense regions; iii) self-gravity in the cloud takes slowly over and local sites of gravitational collapse emerge; iv) global contraction of the cloud starts \citep[for a review, see][]{MK_04, VS_10}. 
\symbolfootnote[0]{$\star$~E-mail: savadd@tu-sofia.bg}

Stars begin to form at stage iii) of the MC evolution, as compressed cloud regions (usually labelled `clumps') evolve and fragment to (prestellar) cores of typical size $0.01 \lesssim l \lesssim 0.1$~pc and densities $n \gtrsim 10^4-10^5$\cc~\citep{BT_07}. Therefore the core mass function (CMF) is considered as a clue to the long-standing problem of the stellar initial mass function (IMF) and its variations in different environments and star formation regions. Indeed, the correspondence between the CMF and the IMF is well established from dust continuum observations of nearby MCs and MC complexes. \citet{TS_98} and \citet{John_ea_01} derive CMFs that could be fitted by a single power-law function of slopes from -1.1 to -1.6 in the mass range $m > 0.5 M_\odot$, in agreement with the Salpeter high-mass slope of the IMF $\Gamma=-1.3$ \citep{Sal55}. Other authors argue for a lognormal \citep{Enoch_ea_08} or a two power-law shape of the CMF that mimics the IMF even better \citep{MAN_98, JMM_06, NW-T_07}. Their results are confirmed by \citet{ALL_07} who used a more reliable approach to obtain core masses from dust extinction measurements toward stellar background. The CMF in the Pipe nebula derived by them is very similar to the IMF but shifted to larger masses by a factor of 4 -- it appears that there is one-to-one correspondence between stars and prestellar cores, assuming star-formation efficiency (SFE) of $\sim$ 25\%. Thus the above-mentioned works suggest that the distinct dense cores are direct progenitors of stars and that the IMF and its characteristic mass are determined by turbulent fragmentation processes in MCs and their fundamental physics \citep{PN_02, Lars_05}. The potential caveats of this approach are discussed by \citet{CKB_07}. 

However, the origin of the CMF from the mass distribution of the initially formed MC clumps is an issue that still needs further elucidation. The clump mass function (ClMF) derived from CO maps exhibits significantly shallower slope $-0.6\gtrsim\Gamma\gtrsim-0.85$ than the IMF and lacks a characteristic mass \citep{Blitz_93, Heit_ea_98, Kramer_ea_98} in contrast to the characteristic mass $M_{\rm ch}\sim 0.5 M_\odot$ of the IMF. Later observational studies allowed for more detailed mapping of MCs. Emission from CO molecules was found to trace lower density cloud regions. The use of other tracers like C$^{18}$O and $^{13}$CO revealed structures with $n \sim 10^4$\cc~but essentially larger ($l\sim0.1-0.5$~pc) than prestellar cores~\citep{Oni_ea_96, Tachi_ea_00, Tachi_ea_02}. Such compact clumps (`dense MC cores') encompass about 10\% of the cloud mass. The derived ClMFs could be represented by a single power-law or a combination of 2 or 3 power-law functions, with a steeper high-mass tail in comparison to the IMF. 

Numerical simulations of clump/core formation caused by supersonic turbulent fragmentation yield a ClMF/CMF characterized by a continuum of slopes, with a high-mass tail steeper than $\Gamma=-1.3$ \citep{BP_ea_06, Schmidt_ea_10}. The velocity dispersion affects significantly the shape of the time-averaged mass distribution -- its characteristic mass decreases and the total number of clumps/cores grows with increasing sonic Mach number ${\cal M}$~\citep{BP_ea_06}. The type of the applied turbulent forcing plays also important role. Compressive forcing produces a shallower high-mass part of the clump/core mass distribution compared to the solenoidal regime \citep{Schmidt_ea_10}. Apparently, a purely turbulent origin of the ClMF is inconsistent with a single power-law behaviour. 

This work presents a novel approach for the derivation of the ClMF as a superposition of mass distributions of clumps, generated by turbulent shocks within a range of spatial scales $0.5 \lesssim L \lesssim 20$~pc. It is based on the study of \citet[][hereafter Paper I]{DVK_11} wherein clumps are defined as condensations formed through a turbulent cascade during the early MC evolution. Our starting point is the power-law relationship $n\propto m^x$ between clump masses and densities which was substantiated in Paper I considering equipartitions relations between various forms of energy. In Section \ref{Phys_description} we list our basic physical assumptions, introduce the parameters of the model, and demonstrate how they fit the observed MC structure. The method to derive the ClMF is described and illustrated in Section~\ref{derive_ClMF}. The results for different choices of clump energy equipartition are presented in Section~\ref{Results}. Section~\ref{Discussion} contains a discussion on our model predictions in view of observational fits of the ClMF and some numerical results. Our conclusions are summarized in Section~\ref{Summary}.

\section{Statistical description of MC structure}  \label{Phys_description} 
A detailed description of the physical framework for this study is given in Paper I. Below we summarize it and illustrate the significance of the free parameters of the model. 

\subsection{Scaling laws} \label{Scaling_laws}
To describe the early evolution of MCs we consider fully developed turbulence creating density structures at any scale in the range $L_{\rm up}\gtrsim L \gtrsim 0.5$~pc through a cascade possibly driven by the very process of cloud formation \citep{KH_10}. The lower limit is close to the transonic scale and is set to provide generated clumps with sizes within the inertial range (see Sect. 4.1 in Paper I). Adopting a largest scale $L_{\rm up}\le20$~pc, we ensure that the gas is mainly molecular and isothermal with typical temperature $T=10$~K. 
 
Turbulent velocity dispersion $u$ and mean mass density $\langle\rho\rangle$ are assumed to scale according to ``Larson's first and second laws'' \citep{Lars_81}:
\begin{equation}
\label{eq_Larson_1}
 u = 1.1\,L^\beta~~~\rm [km\,s^{-1}] 
\end{equation}
\begin{equation}
\label{eq_Larson_2}
 \langle \rho \rangle = 13.6\times10^{-21}\,L^\alpha~~~\rm [g\,cm^{-3}]~, 
\end{equation}
where the coefficient in the second equation is derived adopting a mean molecular mass $\mu=2.4$. The problem whether the power-law indices $\alpha$ and $\beta$ are interdependent is still not resolved \citep[see e.g.][]{BP06}. We choose a fixed value of beta in the range $[0.33, 0.65]$ whereas $\alpha=\alpha(L)$ is derived self-similarly from the assumption of mass-density relationship for clumps generated at a given scale $L$ (see Sect. \ref{n-m_relationship}).

The scaling of the mean magnetic field $B$ is obtained from its relation to the mean mass density. It is widely adopted that $B\propto \langle\rho\rangle^{0.5}$ which is verified from an extensive survey of observational data \citep{Crutch_99}. Thus we get:
\begin{equation}
\label{eq_mag_scaling}
B = 50\,L^{0.5\alpha}~~~\rm [ \mu G]~,
\end{equation}
where the value $B(L=1~{\rm pc})=50~\mu G$ is chosen from the magnetic field scaling in \citet[][Fig. 1 there]{Crutch_99}, assuming a mean density scaling according to ``Larson's second law'' (equation \ref{eq_Larson_2}), with $\alpha\equiv-1$.

\subsection{Clump density distribution}  \label{Lognormal_pdf}
Numerical simulations of supersonic turbulent flows \citep[e.g.][]{Kless_00, LKM_03, P_ea_07, Fed_ea_10} show that the volumetric distribution of (mass) density $\rho$ is described statistically through a standard lognormal probability density function (pdf):
\begin{equation}
\label{eq_pdf}
p(s)\,ds=\frac{1}{\sqrt{2\pi \sigma^2}}\,\exp{\Bigg[-\frac{1}{2}\bigg( \frac{s -s_{\rm peak}}{\sigma}\bigg)^2 \Bigg]}\,ds~,
\end{equation}
where $s\equiv \ln(\rho/\langle\rho\rangle)$, $\sigma$ is the standard deviation (stddev) and $s_{\rm peak}$ is the peak position: 
\begin{equation}
\label{eq_sigma_PDF}
\sigma^2={\rm ln}\,(1+b^2\,{\cal M}^2)~,
\end{equation}
\begin{equation}
\label{eq_max_PDF}
s_{\rm peak}=-\frac{\sigma^2}{2}~.
\end{equation}

The dependence of $\sigma$ on the Mach number determines its scaling through the choice of $\beta$ (equation \ref{eq_Larson_1}). The turbulence forcing parameter $b$ varies in the range $0.2-1.0$, depending on the driving type \citep{Kr_ea_07, FKS_08, Fed_ea_10}. In this paper we adopt the range of values $0.33\le b\le 0.55$ which correspond to transition from purely solenoidal to purely compressive case \citep{FKS_08}.  

In our approach, the clumps generated at a given scale $L$ are represented by a group of statistical objects (`average clump ensemble') resulting from ensemble averaging over the variety of Galactic clouds and cloud complexes (Paper I). The mass density distribution of the ensemble is lognormal, with a peak (equation~\ref{eq_max_PDF}) that corresponds to its most probable member, labelled `typical clump':
\begin{equation}
\label{eq_nc}
 \rho_c=\langle\rho\rangle\,\exp(s_{\rm peak})=\langle\rho\rangle\exp\Big(-\frac{\sigma^2}{2} \Big)
\end{equation}
Then we define the logarithmic density range in the average clump ensemble as $\pm\sigma/2$ from the most probable mass density $\rho_c$:
\begin{equation}
\label{eq_ens_limits}
 \langle\rho\rangle\,\exp\Big(s_{\rm peak}-\frac{\sigma}{2}\Big)\le \rho \le \langle\rho\rangle\,\exp\Big(s_{\rm peak}+\frac{\sigma}{2}\Big)~.
\end{equation}

\subsection{Clump mass-density-size relationship} \label{n-m_relationship}
The basic physical assumption about clumps is the existence of a mass-density relationship: 
\begin{equation}
\label{eq_n-m}
\ln\Big(\frac{\rho}{\rho_0}\Big)=x\,\ln\Big(\frac{m}{m_0}\Big)
\end{equation}
where the power-law index $x$ is assumed to be fixed within a considered clump ensemble, $m$ is clump mass and $\rho_0$ and $m_0$ are arbitrary units of normalization. Adopting the natural presupposition about a statistical relation between clump masses $m$, densities $\rho$ and sizes $l$,
\[ (m/m_0)=(\rho/\rho_0)(l/l_0)^3~, \]
one obtains from equation \ref{eq_n-m} a clump size-density relationship (with size normalization unit $l_0$) as well:
\begin{equation}
\label{eq_n-l}
\ln\Big(\frac{\rho}{\rho_0}\Big)=\frac{3x}{1-x}\,\ln\Big(\frac{l}{l_0}\Big)~.
\end{equation}
Recalling our turbulent scenario of clump formation, the latter relationship should be a self-similar extension of the scaling of density (equation~\ref{eq_Larson_2}), i.e. we adopt: 
\begin{equation}
 \label{eq_dens_scaling}
 \alpha=\frac{3x}{1-x}~.
\end{equation}

The natural density normalization unit is the mean density at the scale in consideration whereas the size normalization unit is chosen to be proportional to the scale size $L$:
\begin{eqnarray}
 \label{eq_norm_rho}
 \rho_0 & \equiv & \langle\rho\rangle \\
 \label{eq_norm_L}
 l_0 & \equiv & \kappa L \\
 \label{eq_norm_m}
 \frac{\rho_0 l_0^3}{m_0} & \propto & \exp\Big(\sigma^2\times\frac{1-x}{x}\Big)
\end{eqnarray}
The last equation is derived in Paper I (equation 19 there). The dimensionless parameter $\kappa$ could be interpreted as mapping resolution of the scale volume. In this work, it is taken to be a constant (Sect.~\ref{MC_structure}); a choice of a scale-dependent functional form does not yield physically meaningful solutions in terms of masses and sizes of the `typical clump'.

\subsection{Equipartition relations}  \label{equipartitions}
As demonstrated in Paper I (see Section 3.1 there), the scale dependence of the exponent $x$ can be derived from equipartition relations between various forms of energy per unit volume\footnote{~Hereafter, the term `energy' is used instead of `energy per unit volume'.}: gravitational $W$, kinetic (turbulent) $E_{\rm kin}$, thermal (internal) $E_{\rm th}$ and magnetic $E_{\rm mag}$. These relations yield a clump size-mass relationship within an `average ensemble' at each scale $L$. Plotting all the ensembles on a single size-mass diagram, one obtains 
a global correlation that can be described by a power law and might be used as a diagnostic tool to verify the models through comparison with simulations (see Fig. 5-8 in Paper I) and observations (Fig.~\ref{fig_l-m_pan}). Best agreement with a numerical study of physics of clumps formed in a weakly magnetized turbulent medium with gravity \citep{Shet_ea_10} was achieved for choices of an equipartition relation that includes both gravitational and kinetic energy. Therefore in this work, we consider the following equipartition relations:

\begin{itemize}
 \item Equipartition of the gravitational vs. kinetic energy:
\begin{equation}
 \label{eq_wkin}  
 |W|\sim f_{\rm gk}E_{\rm kin}~,
\end{equation}
where $f_{\rm gk}$ is a coefficient of proportionality. Such general type of equipartition is expected to hold for structures shaped by turbulence in which gravity gradually takes over. This could happen in regions where turbulence decays locally, or where large-scale flows accumulate material which eventually becomes gravitationally unstable. We adopt a fiducial range $1\le f_{\rm gk}\le 4$ \citep[with an upper limit twice the `virial-like' value $f_{\rm gk}\sim2$; see e.g.][]{VS_ea_07} as could be expected for the early stage of the clump evolution.

 \item Equipartition of the gravitational vs. kinetic and magnetic energy:
 \begin{equation}
  \label{eq_wkin2mag}  
  |W|\sim 2E_{\rm kin}+E_{\rm mag}
 \end{equation}
This case is a form of the `virial-like' equipartition $|W|\sim 2E_{\rm kin}$ taking into account the contribution of magnetic energy which is significant in some dark cloud cores \citep{Crutch_ea_10}. In fact, this equipartition is found as well through numerical simulations \citep{BP_VS_95}.

 \item Equipartition of the gravitational vs. kinetic and thermal energy:
 \begin{equation}
  \label{eq_wkin2th2}  
  |W|\sim 2E_{\rm kin}+2E_{\rm th}
 \end{equation}
This is also a form of the `virial-like' case when the thermal component of the velocity is accounted for. The assumption of this equipartition is relevant at scales where clump sizes are typical for dense cloud cores ($l\lesssim0.3$~pc) and the internal energy becomes comparable to the turbulent one. 
\end{itemize}

\begin{figure*} 
\begin{center}
\includegraphics[width=0.75\textwidth]{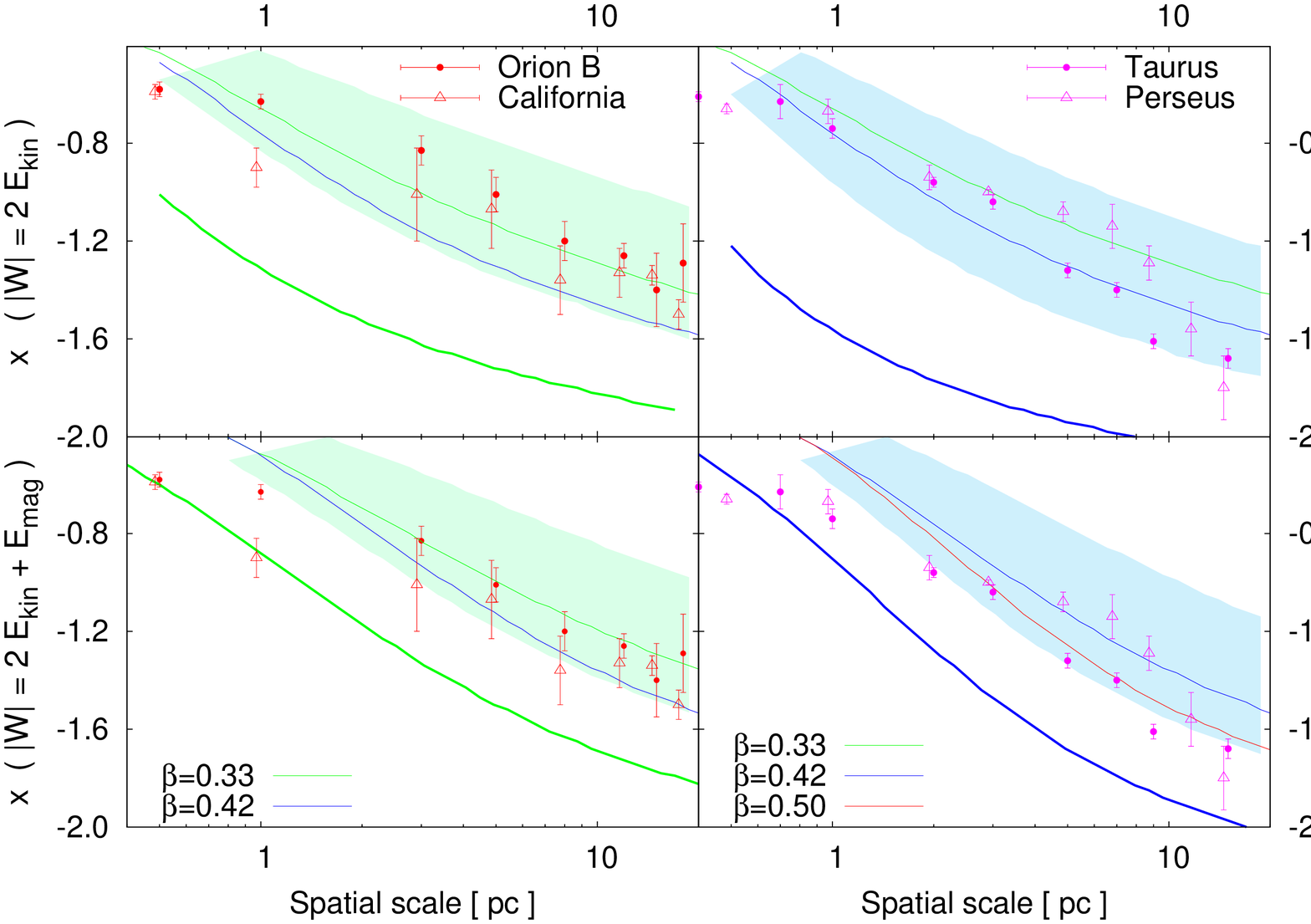}
\vspace{0.4cm}  
\caption{Structure of MCs as traced by the exponent $x(L)$ of the clump mass-density relationship.  Observational estimates \citep{LAL_10} obtained for the clouds Orion B and Taurus (dots) and California and Perseus (triangles) are compared with our model predictions, assuming equipartition relation $|W|\sim2E_{\rm kin}$ (top panels) or $|W|\sim2E_{\rm kin} + E_{\rm mag}$ (bottom panels) and for different sets of the free parameters $(\beta,~b,~\kappa)$. Lines (thin: $b=0.33$; thick: $b=0.55$) denote the choice $\kappa=0.065$ while shaded areas illustrate the effect of varying this parameter ($0.02\le\kappa\le0.10$) when $\beta$ and $b$ are fixed: (\it left) $\beta=0.33$, $b=0.33$; (\it right) $\beta=0.42$, $b=0.33$.}
\label{fig_MCs_kvar_pan}
\end{center}
\end{figure*}

The expressions for the forms of energy of the `typical clump' are given in Appendix \ref{appendix_a}. After substituting them into the equipartition relations (equations \ref{eq_wkin}-\ref{eq_wkin2th2}) and some further transformations, equations for the exponent of the clump mass-density relationship $x$ are derived (Appendix \ref{appendix_b}). The obtained global size-mass correlations are illustrated in Appendix \ref{appendix_c}, together with observational and numerical reference data. The model predictions for cases $|W|\sim2E_{\rm kin}$ (low velocity scaling index $\beta$) and $|W|\sim2E_{\rm kin} + 2E_{\rm th}$ and $|W|\sim4E_{\rm kin}$ ($0.42\le\beta\le0.50$) are in a good consistency with the results from the extensive observational work of \citet{Tachi_ea_02}. We revisit this point again in Sect.~\ref{Discussion} when we comment on the results for ClMF.

\subsection{Modelled and observational structure of MCs}  \label{MC_structure}
We consider an MC as a hierarchical set of spatial scales $L_{\rm up}\gtrsim L \gtrsim 0.5$~pc which is a subset of the inertial range of turbulence. Assuming that the cloud structure is conditioned by balance of energies, it can be described through the solutions $x(L)$ obtained for a chosen equipartition relation. As seen from the equations in Appendix \ref{appendix_b}, those solutions depend on three free parameters: velocity-scaling index $\beta$, turbulent forcing parameter $b$ and mapping resolution $\kappa$ (equation \ref{eq_norm_L}). Hence the sensitivity of $x(L)$ to the choice $(\beta,~b,~\kappa)$ is of critical importance.

\begin{figure*} 
\begin{center}
\includegraphics[width=0.75\textwidth]{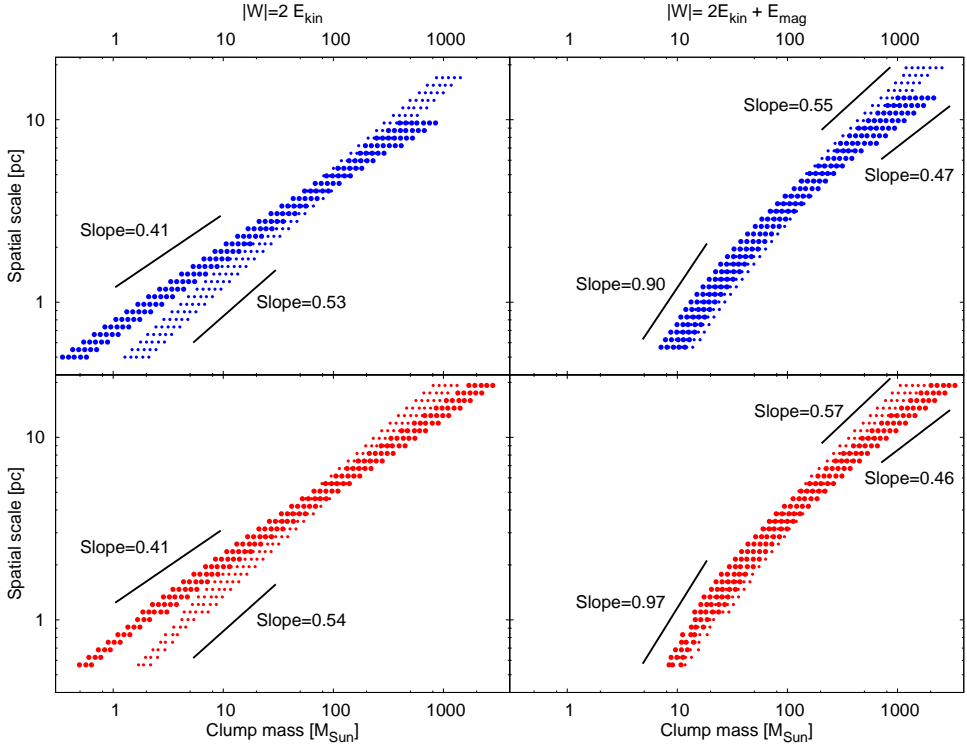}
\vspace{0.3cm}  
\caption{Mass ranges of the `average clump ensemble' vs. the scale of clump generation for different equipartition relations (columns) and chosen turbulent forcing (top: $b=0.55$; bottom: $b=0.33$) and  velocity scaling index (small dots: $\beta=0.33$, larger dots: $\beta=0.50$).} 
\label{fig_m-L_pan}
\end{center}
\end{figure*}

Observational studies of MC structure may be of help to restrict the parameter space that yields plausible solutions. For instance, \citet{LAL_10} derive from extinction maps of several Galactic clouds and cloud complexes a power-law relation between the effective radius $R_{\rm s}=\sqrt{S/\pi}$ of a subregion or a set of subregions with total area $S$ and the mass contained in it/them: $M_{\rm s} \propto R_{\rm s}^{\gamma}$. Since these subregions are not groups of individual clumps but trace the general substructure of a MC, we interprete their effective radii as sizes of spatial scales: $L=2R_{\rm s}$. An `average clump ensemble' is generated at each scale $L$ through a turbulent cascade in the cloud (Sect.~\ref{Lognormal_pdf}). We recall our self-similarity assumption: the scaling law of density within the ensemble is an extension of the scaling law of the mean density (cf. equation \ref{eq_dens_scaling}), i.e. $\langle n \rangle\propto L^{3x/(1-x)}$ . Then the total mass contained within a scale L will be $M\propto\langle n\rangle L^3 \propto L^{3/(1-x)}$. Hence, $\gamma\equiv3/(1-x)$ or: 
\begin{equation}
\label{gamma_x_local}
 x=\frac{\gamma-3}{\gamma}
\end{equation}

Observational estimates of $x(2R_{\rm s})$ obtained in that way from the work of \citet{LAL_10} are compared with our model predictions $x(L)$ in Fig.~\ref{fig_MCs_kvar_pan}. Examples of two types of clouds are selected: with `shallow' (Orion B, California) and with `steep' (Taurus, Perseus) structure in terms of $x(L)$. In general, best fits are obtained for lower values of the velocity scaling index $\beta$ and for mostly solenoidal forcing ($b\gtrsim0.33$). The model predictions are highly sensible to the chosen turbulent forcing - variations of this parameter in the range $0.33\le b\le 0.55$ cause significant shift of the curve $x(L)$ downwards and away from the zone of observational data. On the other hand, variation of $\beta$ affects mainly the shape of this curve, steepening it at larger scales. The value of $\kappa$ should be of order of $10^{-2}$ to achieve a distinction of substructures, significantly smaller than the spatial scale $L$ and significantly larger than the scale of dissipation (Paper I). Evidently, choices of $\kappa$ less than a few percent yield solutions outside the observational range, even for `shallow' clouds like Orion B. Therefore we adopt hereafter a fixed $\kappa=0.065$, a median mapping resolution well below $0.1$. It allows for constructing the ClMF over a limit of confidence of 2-10 solar masses, depending on the assumed equipartition. We comment on that further in Sect.~\ref{Results}.

\section{Derivation of the clump mass function} \label{derive_ClMF}
\subsection{The clump mass-scale diagram}   \label{L-mass-diagram}
The turbulent parameters of each scale $L$: velocity dispersion $u$ (equation \ref{eq_Larson_1}), mean density $\langle \rho \rangle$ (equation \ref{eq_Larson_2}) and forcing parameter $b$ (chosen to be fixed for all scales) determine a lognormal density distribution of clumps $p_{L} (\rho)$ (equation \ref{eq_pdf}). Combining $p_{L} (\rho)$ and the calculated $x (L)$, one obtains a lognormal mass distribution of clumps $p_{L} (m)$, 
\begin{eqnarray}
\label{eq_mass_pdf}
 p_L(m)=\Big(\frac{1}{2 \pi \sigma_m^2}\Big)^{0.5}\,\exp\Bigg[-0.5\bigg(\frac{s_m-s_{m,\,{\rm peak}}}{\sigma_m}\bigg)^2\Bigg]~, \\
 s_{m,\,{\rm peak}}=s_{\rm peak}/x~,~~~~\sigma_m=\sigma/|x|~. \nonumber
\end{eqnarray}

The composite ClMF is derived as a superposition of the clump mass distributions $p_L(m)$, generated within the considered range of scales. A measure of the total number of clumps generated at scale $L$ is the quantity
\begin{equation} 
\label{eq_N_clumps_formula}
 N_{\rm tot}(L)=\frac{1}{\kappa^3}\exp\Big(\sigma^2\times\frac{(x-1)(1-2x)}{2x^2}\Big) 
\end{equation}
as introduced in Paper I (see Eq. 17 there). It serves as a statistical weight for the contribution of a given scale to the composite ClMF. Further, the mass distributions $p_L(m)$ at each scale are presented through discrete sets of weights $\{N_L(m_j)=p_L(m_j)N_{\rm tot}(L),~j=1,...,n \}$ where the range of masses is centred at $s_{m,\,{\rm peak}}(L)$ and its limits are determined from the requirement:
\begin{equation}
 \label{eq_N_clumps_sum}
 N_{\rm tot}(L)=\sum\limits_{j=1}^{n} N_L(m_j)~,~~~\ln\frac{m_{j+1}}{m_j}=h_m~.
\end{equation}

The step $h_m$ must be chosen to be a scale-independent constant, in order to account correctly for the contribution of each scale to a given mass bin of the ClMF. That is evident from diagrams of scale vs. clump mass which are plotted for different equipartition relations and choices of $(\beta,~b)$ in Fig. \ref{fig_m-L_pan}. For $h_m={\rm const}_L$, the distributions $p_L(m)$ correspond to horizontal lines with a fixed linear density of dots. On the other hand, the choice of a constant logarithmic step $h_L=\ln\frac{L_{i+1}}{L_i}$ for the range of scales $L_{\rm down}\le L_i \le L_{\rm up}$ provides a correct counting of clumps as long as a power-law relationship between clump mass and scale of generation holds. Indeed, combining equations \ref{eq_Larson_2}, \ref{eq_nc}, \ref{eq_n-m} and \ref{eq_norm_m}, one gets $m_c\propto m_0 n_c^{1/x} \propto \langle\rho\rangle l_0^3 \exp(-\sigma^2 (3-2x)/2x)\propto L^{3+\alpha}\exp(-\sigma^2 (3-2x)/2x)$. Transforming the index of the exponent in this expression by use of equations \ref{eq_sigma_PDF} and \ref{eq_dens_scaling}, one obtains finally:
\[ m_c \propto L^{3+\alpha}\big(e^{\sigma^2}\big)^{-\frac{3-2x}{2x}} = L^{3+\alpha}\,\Bigg[1+b^2\Big(\frac{1.1}{c_{\rm s}}\Big)^2 L^{2\beta}\Bigg]^{-\frac{\alpha+9}{2\alpha}}\!\!\!, \]
where $c_{\rm s}=0.186$~km/s is the sound velocity at $T=10~\rm K$. The density scaling index $\alpha(L)$ is a smoothly varying function which depends weakly on the chosen equipartition relation. As shown in Fig.~\ref{fig_m-L_pan} (left), an equipartition between gravitational and kinetic energy (equation~\ref{eq_wkin}) yields an approximately constant slope of the mass-scale relation. On the other hand, inclusion of $E_{\rm mag}$ in the equipartition (equation~\ref{eq_wkin2mag}) causes significant change of slope (Fig.~\ref{fig_m-L_pan}, right) at $L\sim 2$~pc: from a steep behaviour at small scales to a value of $\sim 0.50$, quite similar to the one derived from the former relation. This leads to an increased contribution of the small scales to the low- and intermediate-mass ClMF as will be demonstrated in Section~\ref{Discussion}. The mass-scale diagram resulting from equipartition relation between gravitational, kinetic and thermal energy (equation~\ref{eq_wkin2th2}) is quite similar (not plotted) to the one for equation~\ref{eq_wkin}. 

\begin{figure*} 
\begin{center}
\includegraphics[width=1.\textwidth]{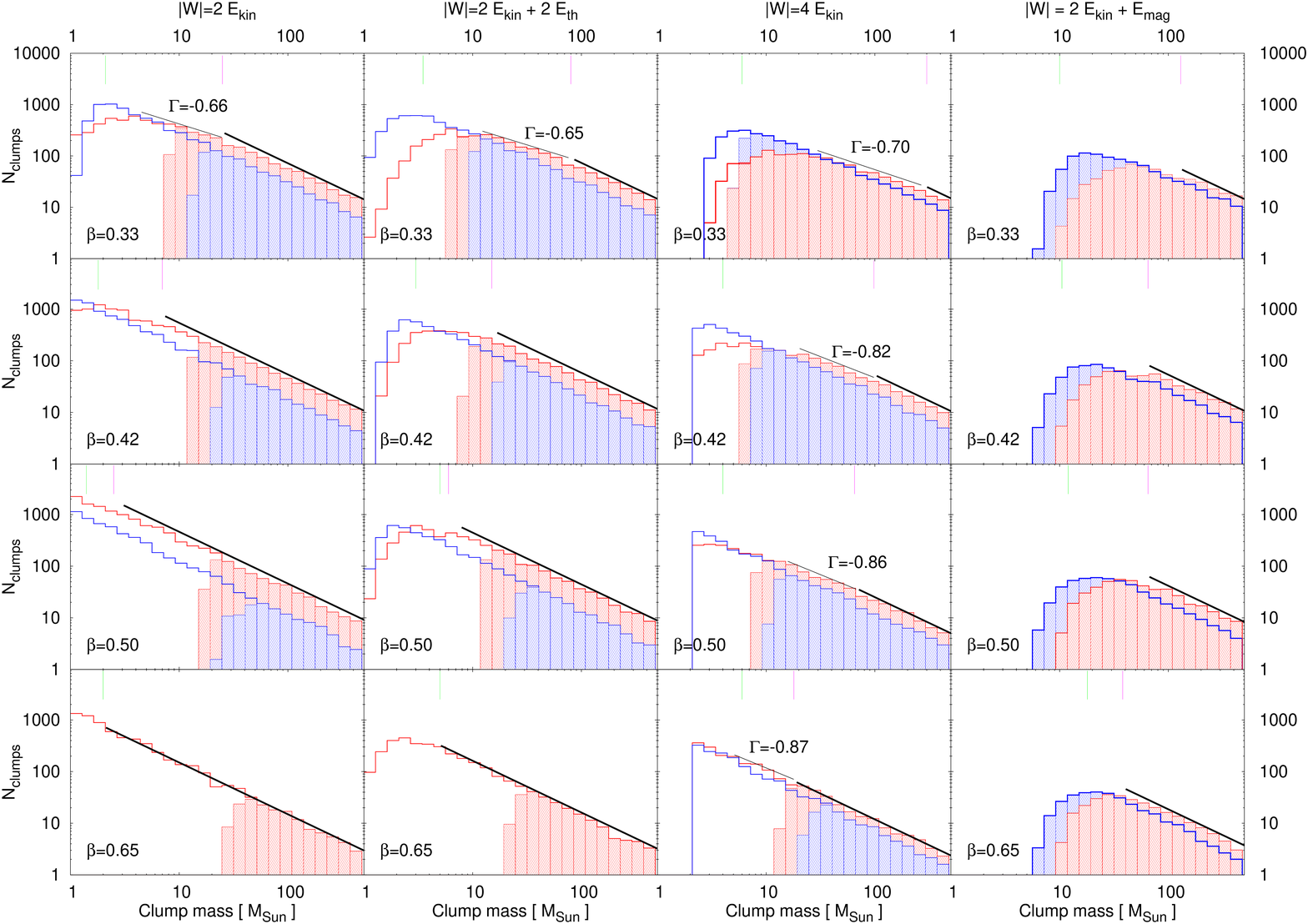}
\vspace{0.4cm}  
\caption{Clump mass function, derived by use of different equipartition relations (columns) and for a fixed velocity scaling index $\beta$ (rows), choosing $b=0.33$ (red) and $b=0.55$ (blue). The ClMF of gravitationally unstable clumps is drawn with hatched areas. The lower mass limit of confidence (green tick mark) and the characteristic mass $M_{\rm ch}$ (violet tick mark) are shown. Black lines denote the predicted slope for fractal clouds $\Gamma=-1$ (thick) and the intermediate-mass ClMF slope(s) for $b=0.33$ (thin).}
\label{fig_cmf_pan}
\end{center}
\end{figure*}

\subsection{Composite ClMF}  \label{Comp_ClMF}
For the derivation of the composite ClMF one must require mass conservation throughout the entire self-similar structure determined by the turbulent cascade process. The largest scale (i.e. the whole cloud) with mass $M(L_{\rm up})$ is to contain ${\cal N}_L$ substructures of mass $M(L)$ at given scale $L<L_{\rm up}$ \citep{Elme_97}. Expressing this equation through the mean density $\langle\rho\rangle=M(L)/L^3$ and using its scaling law (equations \ref{eq_Larson_2}, \ref{eq_dens_scaling}), we get 
\begin{eqnarray}
 \langle\rho\rangle(L_{\rm up})\,L_{\rm up}^3={\cal N}_L\,\langle\rho\rangle(L)\,L^3~, \nonumber \\
 L_{\rm up}^\frac{3x(L_{\rm up})}{1-x(L_{\rm up})}\,L_{\rm up}^3={\cal N}_L\,L^\frac{3x(L)}{1-x(L)}\,L^3 ~,\nonumber
\end{eqnarray}
and hence
\begin{equation}
\label{eq_norm_Ntot}
 {\cal N}_L=L_{\rm up}^\frac{3}{1-x(L_{\rm up})}/L^\frac{3}{1-x(L)}~.
\end{equation}

Note that ${\cal N}_L$ depends essentially on $L$ through $x(L)$. It reflects the fractal structure and can serve as a weight of the total number of clumps at a given scale $N_{\rm tot} (L)$ for derivation of the composite ClMF. Then the ClMF value in a selected mass bin $m^\prime - \Delta m^\prime\le m\le m^\prime + \Delta m^\prime$ is:
\begin{equation}
\label{eq_ClMF}
 F_{\rm ClMF} (m^\prime) = \sum\limits_L {\cal N}_L \sum_{m} N_L(m)~.
\end{equation}

\section{Results} \label{Results}
Brief inspection of the derived ClMFs (Fig.~\ref{fig_cmf_pan}) shows that they can be represented by one or more power-law fits. By analogy with the fitting of the observational IMF \citep[e.g.][]{Kroupa_01}, we define a {\it characteristic mass} $M_{\rm ch}$ as the delimiting value between the high-mass and the intermediate-mass parts of the ClMF. It is calculated as follows: i) choice of a trial value $M_{\rm ch}^{(t)}$; ii) derivation of the slope (least-squares fit) of the high-mass ClMF with lower mass limit $M_{\rm ch}^{(t)}$; iii) the first derived slope ($t=0$) is taken as a standard to compare -- if its value falls within the $3\sigma$ range of the current slope ($t\ge1$), then steps i) and ii) are being repeated. Otherwise, i.e. when the current slope differs significantly, we take $M_{\rm ch}\equiv M_{\rm ch}^{(t)}$ as the characteristic mass and calculate the high-mass ClMF slope for all mass bins $m\gtrsim M_{\rm ch}$.

The ClMFs as derived for different equipartition relations and choices of $(\beta,~b)$, with a fixed largest scale $L_{\rm up}=20$~pc, are plotted in Fig.~\ref{fig_cmf_pan}. Their obvious common feature is the universal power-law shape of the high-mass part $F_{\rm ClMF}\propto m^{\Gamma}$ with $\Gamma\sim-1$ as expected for self-similarly structured fractal clouds \citep{Elme_97, Elme_07}. Variations of $\Gamma$ around this value are found to be minor for the equipartition between gravitational and kinetic energy (column 1 and 3 in Fig.~\ref{fig_cmf_pan}) and slightly larger ($\sigma_{\Gamma}\sim 0.1$) when other forms of energy are included in the equipartition. 

The mass range of the high-mass ClMF depends essentially on the assumed equipartition. In the `virial-like case', without ($|W|\sim 2 E_{\rm kin}$) or with  ($|W|\sim 2 E_{\rm kin} + 2 E_{\rm th}$) inclusion of the thermal energy, it spans from about one to two orders of magnitude as the characteristic mass $M_{\rm ch}$ (violet tick mark) drops from about hundred to several solar masses with increasing the velocity scaling index $\beta$ to $\sim0.50$. If the relative weight of turbulence against gravity is increased or the magnetic energy is included in the equipartition, the same tendency of decreasing of $M_{\rm ch}$ at larger $\beta$ is evident but shifted in the range from few hundred to few tens $M_\odot$. The similarity of the characteristic mass behaviour between the results for $|W|\sim 2 E_{\rm kin}$ and $|W|\sim 2 E_{\rm kin} + 2 E_{\rm th}$ (columns 1-2 in Fig.~\ref{fig_cmf_pan}) or between those for $|W|\sim 4 E_{\rm kin}$ and $|W|\sim 2 E_{\rm kin} + E_{\rm mag}$ (columns 3-4 in Fig.~\ref{fig_cmf_pan}) is remarkable. In the first case, taking into account the contribution of the thermal energy in the clump energy balance leads to increase of $M_{\rm ch}$ with a factor of $3-4$, irrespectively of the turbulent velocity scaling. The results in the second case confirm indirectly (from the equivalence of the considered equipartitions) the existence of a statistical equipartition $2 E_{\rm kin}\sim E_{\rm mag}$ as found by \citet{BP_VS_95}. 

In contrast to the high-mass part, the intermediate-mass ClMF ($M< M_{\rm ch}$) shows up a variety of shapes, corresponding to broad mass ranges from a few to hundred $M_\odot$ (Fig.~\ref{fig_cmf_pan}). When derived from the equipartitions $|W|\sim 2 E_{\rm kin}$ and $|W|=2 E_{\rm kin} + 2 E_{\rm th}$, it can be fitted mainly by a single power-law with slopes $\Gamma\sim-0.65$ as found by some authors for the observational CMF \citep{Blitz_93, Heit_ea_98, Kramer_ea_98}. Robust fitting is possible, however, only in cases with purely solenoidal forcing ($b=0.33$) and for the standard value of velocity scaling $\beta=0.33$ for incompressible turbulence \citep{K_41}. Otherwise it fails due to two main reasons: i) the range between the lower mass limit of confidence (marked with green ticks in Fig.~\ref{fig_cmf_pan}) and $M_{\rm ch}$ is too small; ii) substancial part of the intermediate-mass range is less than 3 mass bins from the turnover of the mass distribution. The choice of equipartition relation $|W|\sim 4 E_{\rm kin}$ or $|W|\sim2 E_{\rm kin} + E_{\rm mag}$ leads to intermediate-mass ClMFs that can be described by two power-law fits. In the first case, the steeper part is easily recognized while fitting of the shallower one is problematic since it contains the turnover of the mass distribution (Fig.~\ref{fig_cmf_pan}, column 3). Their delimiting mass can be obtained in analogical way to the calculation of $M_{\rm ch}$. The increase of the velocity scaling index $\beta$ tends to equalize the slope of the steeper part to that of the high-mass ClMf and to steepen the shallower part. The case $|W|\sim 2E_{\rm kin}+E_{\rm mag}$ is especially interesting because of the flattening of the intermediate-mass ClMF (Fig.~\ref{fig_cmf_pan}, column 4). For turbulence forcing parameter $b=0.33$, the lowest-mass part exhibits even positive slopes. In view of the criteria i) and ii) introduced above, we restrain from fitting the intermediate-mass ClMF in that case.

The choice of a turbulent forcing which is a mixture of solenoidal and compressive mode ($b=0.55$) yields ClMFs of similar shape that do not depend significantly on the assumed equipartition and the velocity scaling index. The tendency with increasing $\beta$ is toward one single power-law (i.e. lack of a characteristic mass) with slope $\Gamma\sim-1$. Again, the case with presence of magnetic energy term is different, showing up $M_{\rm ch}\sim 20-30~M_\odot$. 

The low-mass ClMF is beyond the scope of our consideration due to the lower limit $L_{\rm down}=0.5$~pc of the range of scales (Sect.~\ref{Scaling_laws}) and the relationship between the typical clump mass $m_c$ and the spatial scale of its generation (Sect.~\ref{L-mass-diagram}). At scales $L_{\rm down}\lesssim L\lesssim2$~pc (depending on the chosen equipartition relation), the size of the typical clump approaches the transonic scale which invalidates the assumption of supersonic turbulent fragmentation. The adopted velocity scaling law also fails since $E_{\rm kin}\sim E_{\rm th}$. As shown in Fig.~\ref{fig_cmf_pan}, the confident lower mass limit (green tick mark) of the derived ClMFs varies from a few to about ten solar masses. 

In view of the universal slope of the high-mass ClMF, it is worth to note the effect of varying the largest scale of consideration $L_{\rm up}$. Taking the latter quantity to be the upper limit of the turbulent inertial range, it is proportional to the natal cloud size as demonstrated from simulations \citep{Kr_ea_07, P_ea_07}. Clouds of size $L_{\rm up}\gtrsim12$~pc produce again ClMFs with high-mass slopes $\Gamma\simeq-1$, independent on the chosen equipartition (Fig.~\ref{fig_cmf_Li_var}). Gradual decrease of $L_{\rm up}$ leads to quasi-lognormal shapes without a power law in the high-mass regime. On the other hand, intermediate-mass ClMF with its typical shallower slopes is sustained by the clump generation within a relatively narrow range of scales $0.5\lesssim L \lesssim 5$~pc which upper limit depends on the chosen equipartition relation.  

From the density of a chosen clump (equation \ref{eq_ens_limits}) one can calculate its Jeans mass and hence estimate its gravitational stability. The mass functions of unstable clumps are plotted with hatched boxes in Fig.~\ref{fig_cmf_pan}. The relative fraction of such objects is evidently increasing with the contribution of other forms of energy in the equipartition against gravity -- their lowest masses approach the lower mass limit of confidence in the case $|W|\sim 4 E_{\rm kin}$ while for $|W|\sim 2 E_{\rm kin} + E_{\rm mag}$ all clumps are unstable. Apparently, the mass functions of unstable clumps are not significantly affected by variations of the turbulent forcing parameter $b$. On the other hand, the fraction of unstable clumps decreases -- as expected, -- when $\beta$ is increased.

\section{Discussion}	\label{Discussion}
It seems not surprising to us to obtain a slope $\Gamma=-1$ for the high-mass ClMF, like the typical one for fractal clouds \citep{Elme_97}, in case of equipartition relation between the gravitational and the kinetic energy. In view of the density and velocity scaling laws (Sect.~\ref{Scaling_laws}), turbulence tends to dominate against gravity at larger scales (see Appendix \ref{appendix_b1}) and determines the fractal structure of the cloud and, hence, the high-mass ClMF. The effect on $\Gamma$ of including the thermal energy in the equipartition relation is small since the relative weight of $E_{\rm th}$ decreases with increasing $L$ (see Appendix \ref{appendix_b3}). Inclusion of magnetic energy in the clump energy balance introduces larger, although not essential uncertainties of the high-mass ClMF slope. This could be explained with the strong scale dependence of the magnetic field (equation \ref{eq_mag_scaling}) which results in a steeper relation between the magnetic and kinetic energy terms (see Appendix \ref{appendix_b2}). The effect can be seen in Fig.~\ref{fig_m-L_pan} (right) -- the presence of magnetic field leads to a slow increase of the clump masses with $L$ at small scales while at large scales the clump mass-scale relationship is determined mainly by the kinetic energy. 

The obtained slopes of the intermediate-mass ClMFs deserve special attention because of their (dis)similarities to some observational results on the CMF \citep{Oni_ea_96, Kramer_ea_98, Tachi_ea_02, Kainu_ea_11}. \citet{Kainu_ea_11} derived a single power-law CMF from extinction maps of nearby MCs and obtained a slope $\Gamma=-0.4\pm 0.2$ which is too shallow but possibly affected by clump blending. \citet{Kramer_ea_98} studied several MCs in lines of $^{13}$CO and C$^{18}$O and derived single power-law CMFs of slope $-0.6 \gtrsim \Gamma \gtrsim-0.8$ spanning about 2 orders of magnitude, over a large variety of mass ranges (from small cores up to clouds). Single power-law fits of similar slope result from equipartitions $|W|\sim2 E_{\rm kin}$ and  $|W|\sim2 E_{\rm kin} + 2 E_{\rm th}$ (Fig.~\ref{fig_cmf_obs}, left) and -- in case of larger $b$ and/or $\beta$, -- from $|W|\sim4 E_{\rm kin}$ (Fig.~\ref{fig_cmf_pan}, third column). We note that for Orion B region the choice $|W|\sim2 E_{\rm kin}$ ($\beta=0.33,~b=0.33$) leads not only to a good  agreement between the ClMF and the CMF but also between the predicted MC structure and that derived by \citet{LAL_10} (Fig.~\ref{fig_MCs_kvar_pan}, left top). It should be pointed out that the ClMF mass ranges are severely restricted by the lower mass limit of confidence. In some other cases, the intermediate-mass ClMF is apparently a combination of two power-laws (Fig.~\ref{fig_cmf_obs}, right). The steeper part of it, when derived from equipartition $|W|\sim4 E_{\rm kin}$, for shallow velocity scaling ($0.33\le\beta\le0.42$) and purely solenoidal forcing ($b=0.33$) is again in general agreement with the results of \citet{Kramer_ea_98} while the lower-mass part flattens. Similar behaviour is found by \citet{Tachi_ea_02} from extensive statistics of nearby star-forming regions although these authors obtain essentially steeper slope ($\Gamma=-1.5$) for $m\gtrsim10~M_\odot$. The case $|W|\sim2 E_{\rm kin} + E_{\rm mag}$ yields a combination of positive and flat slopes which could be considered also as a single flat intermediate-mass ClMF (Fig.~\ref{fig_cmf_obs}, bottom right; note the restrictions of the lower mass limit of confidence). The latter fitting is suggested by \citet{Oni_ea_96} for the CMF in Taurus MC ($\Gamma=0.1$). It agrees qualitatively with the intermediate-mass ClMF slopes derived by choosing values of $\beta$ and $b$ that fit best the observed cloud structure for scales $2\lesssim L\lesssim 8$~pc (Fig.~\ref{fig_MCs_kvar_pan}, right bottom). A comparison with Fig.~\ref{fig_m-L_pan} (right bottom) shows that the corresponding mass ranges are $20\lesssim m \lesssim100~M_\odot$ and overlap partially with the mass range of the study of \citet{Oni_ea_96}. To sum up, the best agreement between ClMFs and observational CMFs is apparently achieved in cases $|W|\sim2E_{\rm kin}$ ($0.33\le\beta\le0.42$) and $|W|\sim2E_{\rm kin} + 2E_{\rm th}$ and $|W|\sim4E_{\rm kin}$ ($0.42\le\beta\le0.50$) which is confirmed also by comparison on size-mass diagrams (Fig.~\ref{fig_l-m_pan}). 

Since \citet{Kramer_ea_98} derived CMFs of single clouds, we may speculate that a result from a statistical approach, combining data from more clouds, could be a two- or three power-law CMF. In that aspect, the work of \citet{Tachi_ea_02} is instructive although we still lack an explanation of the steep slopes they derived for $m\gtrsim10~M_\odot$. Their study encompasses objects of sizes, corresponding to our clumps ($0.1\lesssim l \lesssim0.4$~pc), but the variety of their dynamical state is huge: starless, star-forming and cluster-forming cores. In a forthcoming paper, we will address the problem, comparing the predictions of our model for chosen equipartition relation with CMFs derived from molecular-line and dust-continuum data for several Galactic MCs, reflecting various physical conditions. Within this work, we simply point out that different models of cloud structure consistent with the study of \citet{LAL_10} (Fig.~\ref{fig_MCs_kvar_pan};  cf. also Fig. 1 in Paper I) lead to ClMFs which are in a good general agreement with some observational CMFs (Fig.~\ref{fig_cmf_obs}).   

\begin{figure*} 
\begin{center}
\includegraphics[width=0.75\textwidth]{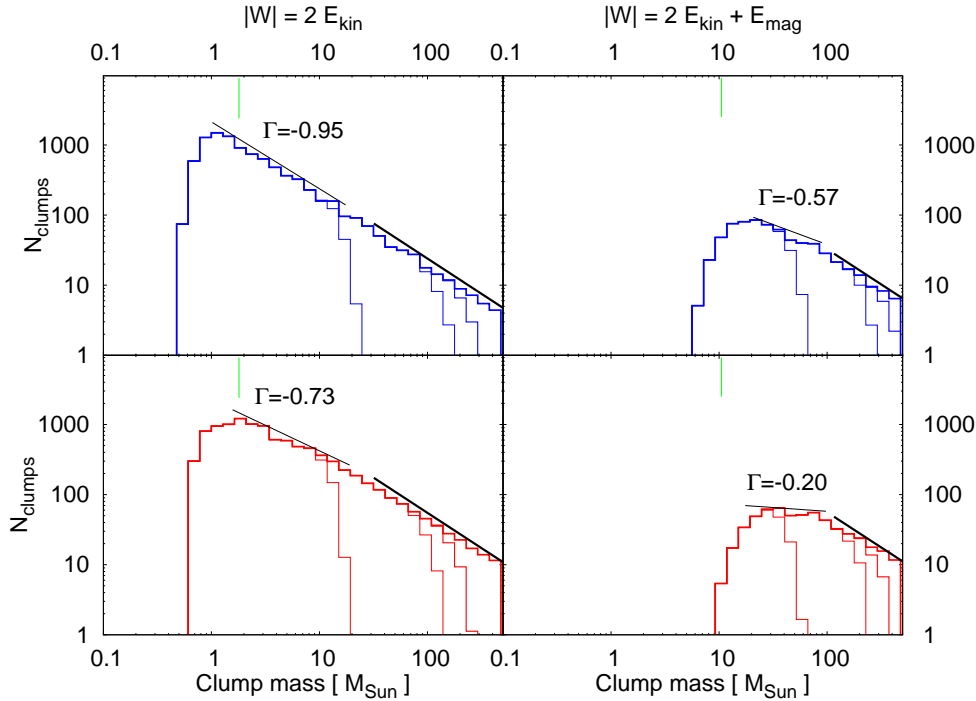}
\vspace{0.4cm}  
\caption{Effect of varying of the upper scale limit $L_{\rm up}=12,~7,~5,~2$~pc (decreasing line-width) on the clump mass function, derived by use of different equipartition relations (columns) and choosing $b=0.33$ (bottom) and $b=0.55$ (top), whereas $\beta=0.42$ is fixed. The slope $\Gamma=-1$ (thick black line) and the lower mass limit of confidence (green tick mark) are plotted for comparison.}
\label{fig_cmf_Li_var} 
\end{center}
\end{figure*}

We obtain quasi-lognormal shapes of the high-mass ClMF only by restricting the upper scale $L_{\rm up}$ of the hierarchy of clumps to very small values (Fig.~\ref{fig_cmf_Li_var}). That result seems to contradict the mass distributions of cores/clumps obtained from numerical simulations of turbulent fragmentation for various different rms Mach numbers \citep{BP_ea_06} and choices of turbulent forcing \citep{Schmidt_ea_10}. However, self-gravity is not included in those simulations while it is a main factor in our approach to derive ClMF from energy equipartition relations. As already commented above, a single power-law high-mass ClMF is a universal feature, resulting from the balance between the gravitational and (mainly) turbulent energy in clumps. Core mass distributions of lognormal shapes and with large widths were derived as well by \citet{Dib_ea_08} from simulations, that include self-gravity. We believe their results are different because their objects are more compact and probably more evolved in comparison with the clumps in our consideration.

Our discussion does not include cases with high turbulent forcing parameter ($1\ge b>0.55$) where the compressive mode provides the main contribution to the turbulent energy \citep{Fed_ea_10}. Our approach is limited in that aspect since the increase of the stddev of the density distribution (equation \ref{eq_sigma_PDF}) often yields unrealistically high typical clump masses $m_c$ in relation to the mass included within the volume of a given scale $L^3$. That constrains the spatial range for a plausible derivation of the ClMF. Such solutions are obtained only for low $\beta\simeq0.33$ and do not differ significantly from those derived for $b=0.55$ (Fig.~\ref{fig_cmf_pan}, upper panels): single power-low with slope about $-1$. Nevertheless, the studied cases for $b=0.55$ are representative enough for the effect of the compressive mode on the ClMF.

\begin{figure*} 
\begin{center}
\includegraphics[width=0.75\textwidth]{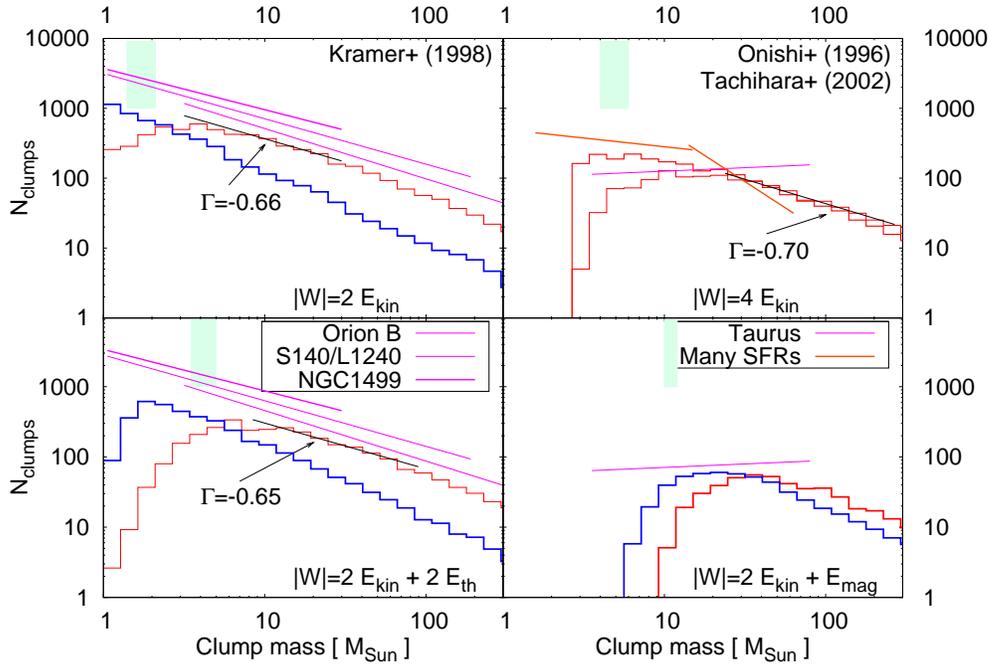}
\vspace{0.4cm}  
\caption{Comparison of derived ClMFs, choosing $b=0.33$ (red) and $b=0.55$ (blue) and $0.33\le\beta\le0.50$ (increasing linewidth), with slopes and mass ranges of some observational CMFs. Variation of the lower mass limit of confidence is shown with green strips.}
\label{fig_cmf_obs}
\end{center}
\end{figure*}

\section{Summary}    \label{Summary}
By use of a statistical approach, we derived mass functions of condensations (clumps) which were formed through a turbulent cascade over a range of spatial scales $L\le20$~pc during the early MC evolution. Clumps are considered within the framework of Paper I: as ensembles of objects in a state of equipartition between gravity and other forms of energy and obeying a power-law mass-density relationship $n\propto m^{x}$. The functional form $x=x(L)$ is determined by the chosen equipartition relation and the free parameters of the model: velocity scaling index $0.33\le\beta\le0.65$ and turbulent forcing parameter $0.33\le b\le0.55$. The clump mass distribution at a fixed scale was obtained from the assumed lognormal density distribution and then the composite clump mass function (ClMF) was derived by superposition of the clump mass distributions generated at the various different scales, assuming self-similar cloud structure. 

The obtained ClMFs for different equipartition relations could be represented by series of power-law functions as intermediate-mass and high-mass parts are distinguished, with a characteristic mass $M_{\rm ch}$ that varies from a few to a few hundred $M_\odot$. The high-mass ClMF can be fitted by a power-law of average slope $\Gamma\simeq-1$, typical for fractal clouds \citep{Elme_97}, with some variations ($\sigma_\Gamma\sim 0.1$) when magnetic energy is included in the energy balance of the clumps. When derived from a `virial-like' equipartition without ($|W|\sim 2 E_{\rm kin}$) or with accounting for the thermal component of the velocity ($|W|\sim 2 E_{\rm kin} + 2 E_{\rm th}$), the intermediate-mass ClMF could be represented by a single power-law of slope $\Gamma\simeq-0.65$, in agreement with some observational clump mass functions (CMFs). Increase of the contribution of turbulent ($|W|\sim 4 E_{\rm kin}$) or magnetic energy ($|W|\sim2 E_{\rm kin} + E_{\rm mag}$) against gravity in the clump energy balance leads to an intermediate-mass ClMF which is a combination of two power-laws, except in the case of large velocity scaling index ($\beta\gtrsim0.50$). The slope of the steeper part varies in a narrow range $-0.7\gtrsim\Gamma\gtrsim-0.9$ depending on the adopted equipartition. The other power law tends to flatten in case of purely solenoidal turbulent forcing ($b=0.33$) and even has a positive slope when the equipartition $|W|\sim 2 E_{\rm kin} + E_{\rm mag}$ is adopted.

Careful comparison with observationally derived mass functions of clumps, considerably larger than low-mass prestellar cores, would demonstrate the ability of our model to match the variety of physical conditions in Galactic star-forming clouds.

{\it Acknowledgement:} T.V. acknowledges support by the {\em Deutsche Forschungsgemeinschaft} (DFG) under grant KL 1358/13-1.

\label{lastpage}
% \newpage
\appendix
\section{Clump energies}
\label{appendix_a}
Below we list the formulae for the terms of energies of the `typical clump' (subscript `c') used in this work.
\begin{itemize}
 \item Gravitational energy:
\begin{equation}
\label{eq_W_clump}
 |W|=z_c\,\frac{3}{5}G\frac{m_c}{l_c/2}\rho_c
\end{equation}
The coefficient $z_c$ accounts for the contribution of the mass outside the clump, typically varying between 1 (vanishing gravitational influence) and 2 (strong gravitational influence from the external cloud) \citep{BP_ea_09}. That range is applicable for scales below the sizes of giant MC and we adopt $z_c=1.5$ (moderate external gravitational influence) in all considered cases. 
 \item Kinetic (turbulent) energy:
\begin{equation}
\label{eq_Ek_clump}
 E_{\rm kin}=\frac{1}{2}\rho_c u_c^2=\frac{1}{2}\rho_c\,u_0^2\,\bigg(\frac{l_c}{1~{\rm pc}}\bigg)^{2\beta} 
\end{equation}
where the typical clump velocity scales like the rms velocity, $u_0=1.1~{\rm km/s}$ (equation \ref{eq_Larson_1}).
 \item Internal (thermal) energy:
\begin{equation}
\label{eq_Ei_clump}
 E_{\rm th}=\frac{3}{2}\frac{\Re}{\mu}\rho_c T~,
\end{equation} 
where $\Re$ is the gas constant. 
 \item Magnetic energy:
\begin{equation}
\label{eq_Em_clump}
 E_{\rm mag}=\frac{B^2(L)}{8\pi}\frac{\rho_c}{\langle\rho\rangle}~,
\end{equation} 
where the magnetic field $B$ scales according to equation \ref{eq_mag_scaling}. 
\end{itemize}

\section{Equipartition functions}\label{appendix_b}
These functions are used to derive the clump mass-density exponent $x$ at a given scale. They are obtained from the equipartition relations (equations \ref{eq_wkin}-\ref{eq_wkin2th2}) by substitutions from the expressions for different clump energies (Appendix \ref{appendix_a}) wherein $l_c$ and $m_c$ are excluded by use of equations \ref{eq_n-l} and \ref{eq_n-m} and the mass normalization unit $m_0$ is expressed from equation \ref{eq_norm_m}.

\subsection{Gravitational vs. kinetic energy}
\label{appendix_b1}
From equation~\ref{eq_wkin}:
\begin{eqnarray}
\label{eq_equipf_wkin}
 Q_{\rm wk}(x) & = & \frac{\pi}{5}z_c G\rho_0 l_0^2 \exp\bigg[-\sigma^2\times\Big(\frac{4x+2}{6x}-\frac{x-1}{x}\Big)\bigg] \nonumber\\ 
& & -f_{\rm gk}\frac{u_0^2}{2} \Big( \frac{l_0}{1~\rm pc}\Big)^{2\beta}\!\!\exp\Big(-\sigma^2\times\frac{2\beta+(3-2\beta)x}{6x}\Big)
\end{eqnarray}

\subsection{Gravitational vs. kinetic and magnetic energy}
\label{appendix_b2}
From equation~\ref{eq_wkin2mag}:
\begin{eqnarray}
\label{eq_equipf_wkin2mag}
 Q_{\rm wkmag}(x) & = & \frac{\pi}{5}z_c G\rho_0 l_0^2 \exp\bigg[-\sigma^2\times\Big(\frac{4x+2}{6x}-\frac{x-1}{x}\Big)\bigg] \nonumber\\
 & & -u_0^2 \Big( \frac{l_0}{1~\rm pc}\Big)^{2\beta}\exp\Big(-\sigma^2\times\frac{2\beta+(3-2\beta)x}{6x}\Big) \nonumber\\
 & & - \frac{B^2}{8\pi\rho_0}\exp\Big(-\frac{\sigma^2}{2}\Big)
\end{eqnarray}

\subsection{Gravitational vs. kinetic and thermal energy}
\label{appendix_b3}
From equation~\ref{eq_wkin2th2}: 
\begin{eqnarray}
\label{eq_equipf_wkin2th2}
  Q_{\rm wkth}(x) & = & \frac{\pi}{5}z_c G\rho_0 l_0^2 \exp\bigg[-\sigma^2\times\Big(\frac{4x+2}{6x}-\frac{x-1}{x}\Big)\bigg] \nonumber\\ 
& & - u_0^2 \Big( \frac{l_0}{1~\rm pc}\Big)^{2\beta}\exp\Big(-\sigma^2\times\frac{2\beta+(3-2\beta)x}{6x}\Big) \nonumber\\
& & - 3 \exp\Big(-\frac{\sigma^2}{2}\Big)\frac{\Re T}{\mu}
\end{eqnarray}

\section{Size-mass diagrams}\label{appendix_c}
\begin{figure*} 
\begin{center}
\includegraphics[width=1.\textwidth]{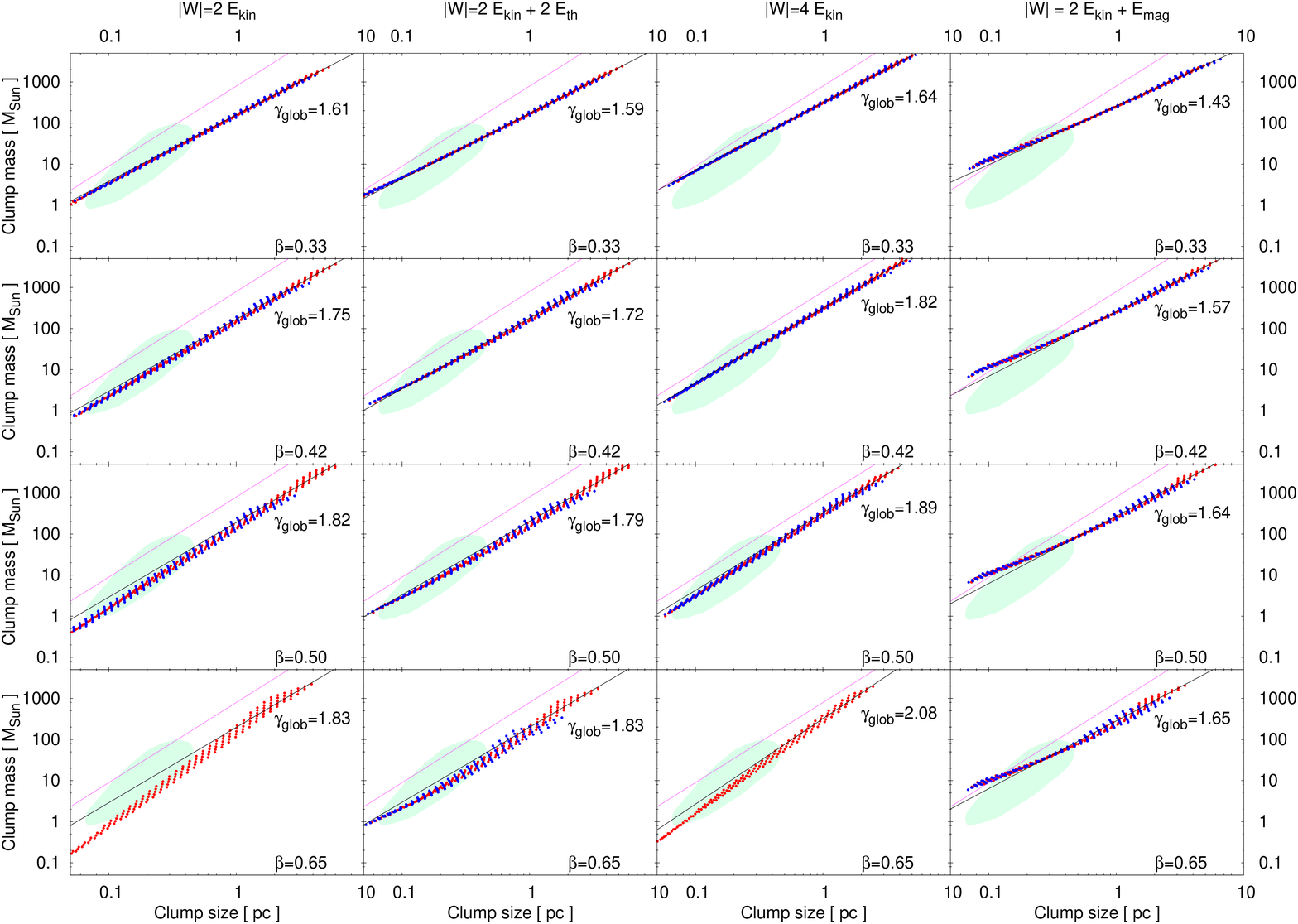}
\vspace{0.4cm}  
\caption{Size-mass diagrams, composed for different equipartition relations (columns) and for a fixed velocity scaling index $\beta$ (rows), choosing $b=0.33$ (red) and $b=0.55$ (blue). The derived slope for each chosen method and parameter set is given (solid black line). The region of the cloud cores studied by \citet{Tachi_ea_02} (green hatched areas) and the slope $-1.95$ obtained for clumps from the simulation of \citet{Shet_ea_10} (violet line) are plotted for comparison.}
\label{fig_l-m_pan}
\end{center}
\end{figure*}

Plotted on size-mass diagrams, the `average clump ensembles' generated at each scale $L$ exhibit global power-law correlations with slopes $\gamma_{\rm glob}$. In Fig.~\ref{fig_l-m_pan}, they are juxtaposed with reference data from observations of cloud cores \citep{Tachi_ea_02} and simulations of cloud clumps formed in a weakly magnetized turbulent medium with gravity \citep{Shet_ea_10}. In the latter case, clumps have been delinated analogically to an observational study of MC structures, imposing a cut-off level on the column density map.

\end{document}